\documentclass[twocolumn]{aastex631}
\usepackage{graphicx}

\shorttitle{Similarity between JWST's compact EROs in cosmic dawn and BluDOGs known in cosmic noon}
\shortauthors{Noboriguchi et al.}

\begin{document}

\title{Similarity between compact extremely red objects discovered with JWST in cosmic dawn and blue-excess dust-obscured galaxies known in cosmic noon}

\correspondingauthor{Akatoki Noboriguchi}
\email{akatoki@shinshu-u.ac.jp}
\correspondingauthor{Akio K. Inoue}
\email{akinoue@aoni.waseda.jp}

\author[0000-0002-5197-8944]{Akatoki Noboriguchi}
\affiliation{Center for General Education, Shinshu University, 3-1-1 Asahi, Matsumoto, Nagano 390-8621, Japan}

\author[0000-0002-7779-8677]{Akio K. Inoue}
\affiliation{Waseda Research Institute for Science and Engineering, Faculty of Science and Engineering, Waseda University, 3-4-1, Okubo, Shinjuku, Tokyo 169-8555, Japan}
\affiliation{Department of Physics, School of Advanced Science and Engineering, Faculty of Science and Engineering, Waseda University, 3-4-1, Okubo, Shinjuku, Tokyo 169-8555, Japan}
\author[0000-0002-7402-5441]{Tohru Nagao}
\affiliation{Research Center for Space and Cosmic Evolution, Ehime University, 2-5 Bunkyo-cho, Matsuyama, Ehime 790-8577, Japan}
\author[0000-0002-3531-7863]{Yoshiki Toba}
\affiliation{National Astronomical Observatory of Japan, 2-21-1 Osawa, Mitaka, Tokyo 181-8588, Japan}
\affiliation{Academia Sinica Institute of Astronomy and Astrophysics, 11F of Astronomy-Mathematics Building, AS/NTU, No.1, Section 4, Roosevelt Road, Taipei 10617, Taiwan}
\affiliation{Research Center for Space and Cosmic Evolution, Ehime University, 2-5 Bunkyo-cho, Matsuyama, Ehime 790-8577, Japan}
\author[0000-0002-5464-9943]{Toru Misawa}
\affiliation{Center for General Education, Shinshu University, 3-1-1 Asahi, Matsumoto, Nagano 390-8621, Japan}

\begin{abstract}
Spatially compact objects with extremely red color in the rest-frame optical to near-infrared (0.4--1 ${\rm \mu m}$) and blue color in the rest-frame ultraviolet (UV; 0.2--0.4 ${\rm \mu m}$) have been discovered at $5 < z < 9$ using the James Webb Space Telescope (JWST).
These extremely red objects (JWST-EROs) exhibit spectral energy distributions (SEDs) that are difficult to explain using a single component of either star-forming galaxies or quasars, leading to two-component models in which the blue UV and extremely red optical are explained using less-dusty and dusty spectra of galaxies or quasars, respectively.
Here, we report the remarkable similarity in SEDs between JWST-EROs and blue-excess dust-obscured galaxies (BluDOGs) identified at $2 < z < 3$.
BluDOGs are a population of active galactic nuclei (AGNs) with blackhole masses of $\sim10^{8-9}$ M$_\odot$, which are one order of magnitude larger than those in some JWST-EROs.
The Eddington ratios of BluDOGs are one or higher, whereas those of JWST-EROs are in the range of 0.1--1.
Therefore, JWST-EROs are less massive, less active, and more common counterparts in higher-$z$ of BluDOGs in cosmic noon.
Conversely, JWST-EROs have a significantly higher fraction of those with blue-excess than DOGs.
We present the average UV spectra of BluDOGs as a comparison to JWST-EROs and discuss a coherent evolutionary scenario for dusty AGN populations.
\end{abstract}


\section{Introduction} \label{sec:intro}

The James Webb Space Telescope (JWST) has opened up an amazing new window to the very early Universe with a near-infrared (NIR) camera (NIRCam), NIR Spectrograph (NIRSpec), and mid-infrared (MIR) instrument (MIRI).
NIR and MIR wavelengths are important for investigating high-$z$ objects, as emission lines in the rest-frame ultraviolet (UV) and optical are shifted to NIR and MIR.
Recently, spatially compact and extremely red objects (EROs) have been discovered in the observational data of the JWST \citep{2023arXiv230200012K, 2023arXiv230412347A, 2023arXiv230514418B, 2023arXiv230605448M, 2023arXiv230607320L,2023arXiv230805735F, 2023ApJ...952..142F,2023arXiv230811610K}.
We refer to these objects as JWST-EROs in this Letter. 
The JWST-EROs exhibit a red color between 2.77 and 4.44 ${\rm \mu m}$-bands ($(F277W-F444W)_{\rm AB} > 1.5$) and a blue color between 1.50 and 2.00 ${\rm \mu m}$-bands ($(F150W-F200W)_{\rm AB} \sim 0$) \citep{2023arXiv230514418B}.
Given the spectroscopic redshifts ($z_{\rm spec}$) or photometric redshifts ($z_{\rm photo}$) of the JWST-EROs, spectral energy distributions (SEDs) are characterized by a peculiar combination of the extremely red color in the rest-frame 0.4--1 ${\rm \mu m}$ and blue color in the rest-frame 0.2--0.4 ${\rm \mu m}$.
Such SEDs are difficult to explain with a single population of galaxies or quasars, but they can be explained with composites of two components of less-dusty galaxies/quasars and dusty galaxies/quasars \citep{2023arXiv230200012K, 2023arXiv230412347A, 2023arXiv230514418B, 2023arXiv230607320L}.
The spatial compactness of the JWST-EROs suggests potential active galactic nuclei (AGNs) \citep{2023arXiv230412347A, 2023arXiv230514418B, 2023arXiv230607320L}.
Some JWST-EROs exhibit broad emission lines in their spectra, indicating that they are AGNs \citep{2023arXiv230200012K, 2023arXiv230605448M, 2023arXiv230805735F, 2023arXiv230811610K}.

There is a population of dusty AGNs called dust-obscured galaxies (DOGs), which are thought to be in a transition phase between dusty star formation and dusty AGNs after a gas-rich major merger event \citep{2008ApJ...677..943D}.
DOGs are AGNs selected by a color between observed-frame optical and MIR.
\cite{2015PASJ...67...86T,2017ApJ...835...36T} selected DOGs from Subaru Hyper Suprime Cam (HSC; \citealt{2018PASJ...70S...1M})-Subaru Strategic Program (SSP; \citealt{2018PASJ...70S...8A}) data and {\textit{Wide-field Infrared Survey Explorer}} ({\textit{WISE}}; \citealt{2010AJ....140.1868W}) data using the color criterion ($(i-W4)_{\rm AB} \geq 7.0$, where $i$ and $W4$ denotes the magnitudes of $i$- and $W4$-bands) \citep[see also][]{2016ApJ...820...46T}.
The extremely red color between optical and MIR is explained by heavy dust reddening in UV/optical and re-emission in MIR from the dusty torus surrounding the nucleus.

Most DOGs exhibit a simple red SED described by a power law \citep[see e.g.,][]{2020ApJ...888....8T,2020ApJ...889...76T}.
However, \cite{2019ApJ...876..132N} found eight DOGs with blue-excess in optical bands (blue-excess DOGs; BluDOGs) using an observed-frame optical slope ($\alpha_{\rm opt} < 0.4$, where $\alpha_{\rm opt}$ denotes the observed-frame optical spectral index of the power law fitted to the HSC $g$-, $r$-, $i$-, $z$-, and $y$-band fluxes, $f_{\nu}\propto\lambda^{\alpha_{\rm opt}}$).
After spectroscopic follow-up observations, \cite{2022ApJ...941..195N} reported that BluDOGs are broad-line AGNs at $2<z<3$ and the origin of the blue-excess are blue continuum and large equivalent widths (EWs) of the broad emission lines.
In addition, C~{\sc iv} lines exhibit a blue tail, suggesting that BluDOGs have nuclear outflows \citep{2022ApJ...941..195N}.
The Eddington ratio ($\lambda_{\rm Edd} = L_{\rm bol}/L_{\rm Edd}$, where $L_{\rm bol}$ and $L_{\rm Edd}$ denote the bolometric luminosity and the Eddington luminosity, respectively) of BluDOGs is greater than 1, i.e., it is in a super Eddington phase.
Therefore, BluDOGs are likely to be in a transition phase between dusty AGN and optically thin quasar phases \citep{2022ApJ...941..195N}.

In this Letter, we present the remarkable similarity in SEDs between the JWST-EROs at $z>5$ and BluDOGs at $2<z<3$.
This Letter is structured as follows.
We describe the samples of the JWST-EROs and BluDOGs in Section~\ref{sec:sample}.
In Section~\ref{sec:results}, we compare the SEDs and physical parameters between JWST-EROs and BluDOGs.
In Section~\ref{sec:sd}, we discuss the number densities and excess UV emission of JWST-EROs and BluDOGs and present a possible evolutionary scenario of dusty AGNs to explain the similarity between JWST-EROs and BluDOGs.
Throughout this Letter, the adopted cosmology is a flat universe with $H_0 = 70$ km s$^{-1}$ Mpc$^{-1}$, $\Omega_M = 0.3$, and $\Omega_\Lambda = 0.7$.
Unless otherwise stated, all magnitudes refer to the AB system \citep{1983ApJ...266..713O}.

\section{Sample} \label{sec:sample}

In this Letter, we use the JWST-ERO samples taken from \cite{2023arXiv230514418B} and \cite{2023arXiv230605448M} and the BluDOG sample taken from \cite{2022ApJ...941..195N}.

\cite{2023arXiv230514418B} selected 37 JWST-EROs in CEERS fields \citep{2023ApJ...946L..13F} based on the single color criterion $(F277W-F444W)_{\rm AB} > 1.5$. 
These EROs have an average magnitude of $F444W=25.9$ AB mag and $5<z_{\rm photo}<9$.
Surprisingly, their color $(F150W-F277W)_{\rm AB} \sim 0$ indicates a flat slope in the rest-frame UV in contrast to their very red rest-frame optical color. 
In NIRCam images, these EROs are generally unresolved, point-like sources.
\cite{2023arXiv230514418B} have also reported that among 37 JWST-EROs, four objects are found in the MIRI imaging area of CEERS and that these objects are detected as consistent with an extrapolation from $F444W$ with the red slope.
Another set of four EROs is found in the NIRSpec targets of CEERS, and they exhibit clear emission lines, securing robust spectroscopic redshifts.
One of them is the broad-line AGN at $z_{\rm spec}=5.62$ reported in \cite{2023arXiv230200012K}.
Because \cite{2023arXiv230514418B} presented individual SEDs of the eight JWST-EROs in addition to two stacked SEDs of the 37 JWST-EROs divided into two groups depending on their photometric redshifts, we adopt these SEDs as typical SEDs of JWST-EROs in this Letter.

\cite{2023arXiv230605448M} identified 20 broad-line ($>1000$ km s$^{-1}$) H$\alpha$ emitters at $z\sim5$ from the wide-field slitless spectroscopy data of the EIGER \citep{2023ApJ...950...66K} and FRESCO \citep{2023arXiv230402026O} surveys.
These objects are generally spatially compact point-like sources, except in some cases with faint companions.
\cite{2023arXiv230605448M} concluded that these H$\alpha$ emitters are AGNs because of their broad emission line and compact morphology.
The blackhole masses and bolometric luminosities of the H$\alpha$ emitters are estimated as log$_{10} (M_{\rm BH}/M_\odot)=6.9$--$8.6$ and $L_{\rm bol}=5.0$--$65.8\times10^{44}$ erg s$^{-1}$, respectively, from the H$\alpha$ line widths and luminosities \citep{2023arXiv230605448M}.
Although \cite{2023arXiv230605448M} did not adopt any color criteria for the selection, the colors of many H$\alpha$ emitters are similar to those of JWST-EROs of \cite{2023arXiv230514418B}, i.e., $(F210M-F444W)_{\rm AB} > 1.5$ or $(F200W-F356W)_{\rm AB} > 0.8$, and $(F182M-F210)_{\rm AB} \sim 0.0$ or $0.0 < (F115W-F200W)_{\rm AB} < 1.2$ (see Figure 2 in \citealt{2023arXiv230605448M}).

The BluDOG sample consists of eight objects selected by \cite{2019ApJ...876..132N}.
\cite{2022ApJ...941..195N} conducted spectroscopic observations for four brightest BluDOGs ($r_{\rm AB} < 23$) among them.
The spectroscopic redshifts are between 2.2 and 3.3.
\cite{2022ApJ...941..195N} also estimated the blackhole masses as $1.1\times10^8<M_{\rm BH}/M_{\odot}<5.5\times10^8$ based on the C~{\sc iv} emission lines using the calibration formula of \cite{2006ApJ...641..689V}.
The bolometric luminosities of the BluDOGs are estimated using SED fitting, and their inferred Eddington ratios are greater than 1 ($1.1<\lambda_{\rm Edd}<3.8$;  \citealt{2022ApJ...941..195N}).

\section{Results} \label{sec:results}

\subsection{Comparison of SEDs of JWST-EROs and BluDOGs\label{ss:com_SED}}

\begin{figure*}
\includegraphics[width=18cm]{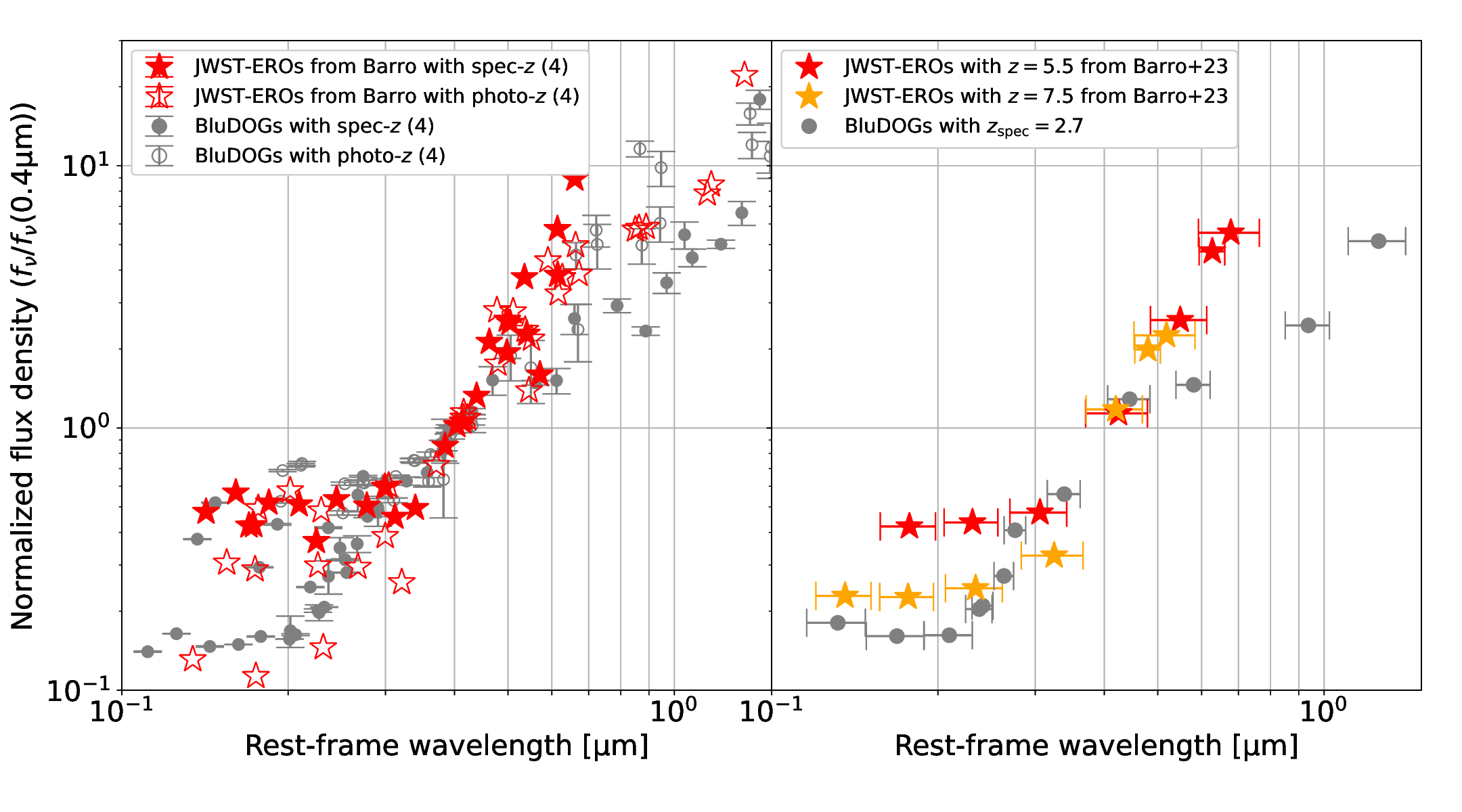}
\caption{Left panel: Comparison of individual SEDs of eight EROs (red stars) at $5<z<8$ found with JWST \citep{2023arXiv230514418B} and eight BluDOGs (gray circles) at $2<z<3$ \citep{2022ApJ...941..195N}.
The filled and open symbols represent objects with spec-$z$ and photo-$z$, respectively. 
Right panel: Comparison of average SEDs of JWST-EROs and BluDOGs.
The red and orange stars represent the averages of JWST-EROs with $z_{\rm photo}$ = 5.5 and 7.5, respectively. 
The gray filled circles represent the average of BluDOGs with spec-$z$. 
Both galaxy populations share the characteristic bimodal colors: blue UV and extremely red optical colors.
All flux densities are normalized by that at the rest-frame 0.4 $\mu$m.
\label{fig:SED}}
\end{figure*}

First, we compare the SEDs of JWST-EROs \citep{2023arXiv230514418B} and those of BluDOGs \citep{2022ApJ...941..195N} in Figure~\ref{fig:SED}. 
There is a remarkable similarity in SEDs between the two galaxy populations; the SEDs are characterized by an extremely red optical color and a flat blue UV color.
This characteristic ``bimodal-color'' SED in UV and optical is seen both in comparisons of individual SEDs (left panel of Fig.~\ref{fig:SED}) and of average SEDs (right panel).
Although the overall SEDs are very similar, JWST-EROs exhibit even redder optical colors than BluDOGs, as seen in the comparison of average SEDs (right panel).
This could be caused by photometric excesses found in some cases at the rest-frame 0.6--0.7 $\mu$m in individual SEDs (left panel), which can be caused by a strong H$\alpha$ emission line, as shown in \cite{2023arXiv230514418B}.
Another difference is the wavelength of the ``spectral break'' between UV and optical.
In the average SEDs, JWST-EROs exhibit a break around the rest-frame 0.3 $\mu$m, whereas BluDOGs exhibit a break around 0.2 $\mu$m.
However, the break wavelengths are dispersed in individual SEDs.
In summary, the SEDs of JWST-EROs and BluDOGs are remarkably similar, strongly suggesting their physical connection.

\subsection{UV and Optical Spectral Slopes\label{ss:beta-uv-opt}}

\begin{figure}
\includegraphics[width=8.5cm]{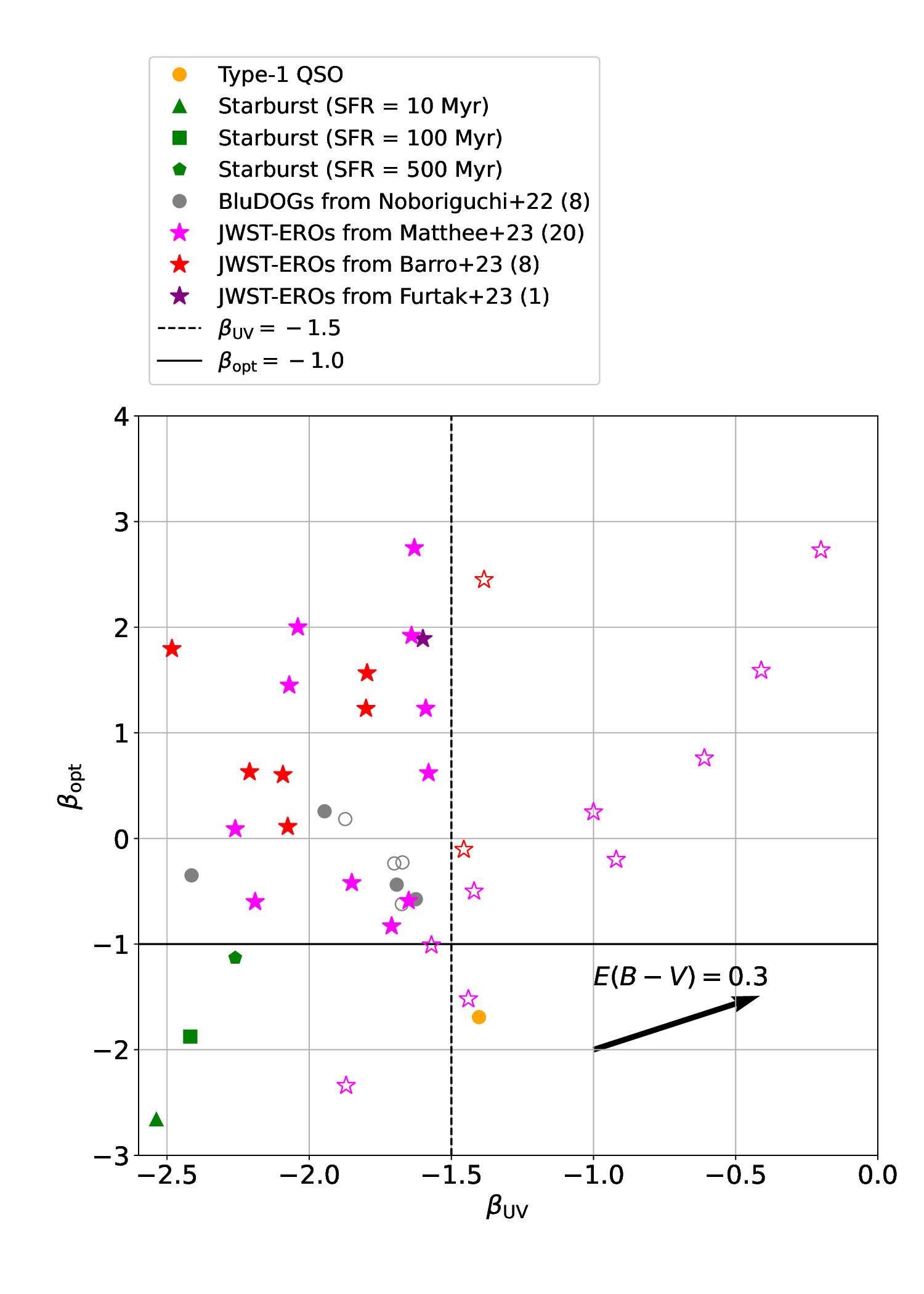}
\caption{$\beta_{\rm UV}$ vs.\ $\beta_{\rm opt}$ diagram. 
The magenta, red, and purple stars represent JWST-EROs from \cite{2023arXiv230605448M}, \cite{2023arXiv230514418B}, and \cite{2023ApJ...952..142F}, respectively.
The filled stars have $\beta_{\rm UV}<-1.5$ and $\beta_{\rm opt}>-1.0$, and the open stars have $\beta_{\rm UV}>-1.5$ or $\beta_{\rm opt}<-1.0$.
The gray filled and open circles represent BluDOGs with spec-$z$ and photo-$z$ from \cite{2022ApJ...941..195N}, respectively. 
The orange circle represents Type-1 QSO from the SWIRE template library \citep{2007ApJ...663...81P}.
The green triangle, square, and pentagon plots denote templates of star-forming galaxies with metallicity of 0.2 Solar value and a constant SFR with durations of 10, 100, and 500 Myr, respectively \citep{2011MNRAS.415.2920I}.
The black solid and dashed lines denote $\beta_{\rm opt} = -1.0$ and $\beta_{\rm UV} = -1.5$, respectively.
The black arrow represents a reddening vector with $E(B-V)=0.3$ for the Calzetti law \citep{2000ApJ...533..682C}. 
\label{fig:beta}}
\end{figure}

Second, we compare the UV and optical spectral slopes of the samples in Figure~\ref{fig:beta}, where $\beta_{\rm UV}$ and $\beta_{\rm opt}$ are defined as the spectral indices in $f_\lambda\propto\lambda^\beta$ for the rest-frame UV ($\sim0.2$~$\mu$m) and optical ($\sim0.5$~$\mu$m), respectively.
\cite{2023arXiv230605448M} reported the measurements of $\beta_{\rm UV}$ and $\beta_{\rm opt}$ of the broad-line H$\alpha$ emitters by applying power-law fitting to photometric data at observed wavelengths of 1--2 ${\rm \mu m}$ and 2--4 ${\rm \mu m}$, respectively.
These wavelength ranges correspond to the rest-frame UV and optical.
For the eight individual EROs from \cite{2023arXiv230514418B}, we have measured $\beta_{\rm UV}$ and $\beta_{\rm opt}$ using the photometric data in the same wavelength ranges as those used in \cite{2023arXiv230605448M}.
In addition, we found $\beta_{\rm UV}$ and $\beta_{\rm opt}$ values of one JWST-ERO reported by \cite{2023arXiv230805735F}.
As depicted in Figure~\ref{fig:beta}, these JWST samples share the distribution in the $\beta_{\rm UV}$--$\beta_{\rm opt}$ diagram, indicating the similarity of the overall SEDs.
That is, the optical color is very red ($\beta_{\rm opt}>-1$), and the UV color is blue ($\beta_{\rm UV}\sim -2$), with few exceptions. 
Therefore, the broad-line H$\alpha$ emitters from \cite{2023arXiv230605448M} are also called JWST-EROs in this Letter.

The BluDOGs are also plotted on the $\beta_{\rm UV}$--$\beta_{\rm opt}$ diagram of Figure~\ref{fig:beta}. 
$\beta_{\rm UV}$ of BluDOGs are calculated by applying power-law fitting to $grizy$-bands of Subaru/HSC \citep{2022ApJ...941..195N}.

We also calculate the $\beta_{\rm UV}$ and $\beta_{\rm opt}$ for Type-1 QSO and star-forming galaxy (SFG) templates as comparison data. 
The Type-1 QSO template is taken from the SWIRE template library \citep{2007ApJ...663...81P}, and the SFG templates are the cases with constant star formations of 10, 100, and 500 Myr \citep{2011MNRAS.415.2920I}.
By defining the wavelength ranges of $\beta_{\rm UV}$ and $\beta_{\rm opt}$ as the rest-frame 1500--3000 and 3000--6000 \AA, respectively, we calculate $\beta_{\rm UV}$ and $\beta_{\rm opt}$ using power-law fitting.
The $\beta_{\rm UV}$ and $\beta_{\rm opt}$ of the Type-1 QSO are $-1.40$ and $-1.69$, respectively.
The SFG templates of the ages $=$ 10, 100, and 500 Myr have ($\beta_{\rm UV}$, $\beta_{\rm opt}$) = ($-2.54, -2.66$), ($-2.42, -1.88$), and ($-2.26, -1.13$), respectively.

Figure~\ref{fig:beta} shows the resultant $\beta_{\rm UV}$--$\beta_{\rm opt}$ diagram.
We find that the UV and optical colors of the JWST samples are similar, blue in UV but extremely red in optical, while some objects are very red in UV and a few objects are blue in both UV and optical.
The BluDOGs are found near the center of the distribution of the JWST samples, indicating the similarity of the overall SEDs between these galaxy populations.
Conversely, SFGs are significantly bluer both in UV and optical, and Type-1 QSO is also significantly bluer in optical but similar in color or slightly redder in UV.
Although arbitrary, we select objects with $\beta_{\rm UV}<-1.5$ (bluer than the Type-1 QSO) and $\beta_{\rm opt}>-1.0$ (redder than the SFGs) as BluDOG-like objects, and 18 of 29 JWST-EROs satisfy these criteria.
These BluDOG-like objects are located in different areas in the $\beta_{\rm UV}$--$\beta_{\rm opt}$ diagram from the SFGs and Type-1 QSO, which are common objects in $z>3$.
Even if we consider SFGs and Type-1 QSOs with dust reddening, it is difficult to reproduce their red optical and blue UV colors simultaneously, as previously reported (e.g., \citealt{2023arXiv230412347A, 2023arXiv230514418B, 2023arXiv230607320L}).

\subsection{Distribution in $M_{\rm BH}$ vs.\ $L_{\rm bol}$ Diagram\label{ss:com_MBH_Lbol}}

\begin{figure}
\includegraphics[width=8.5cm]{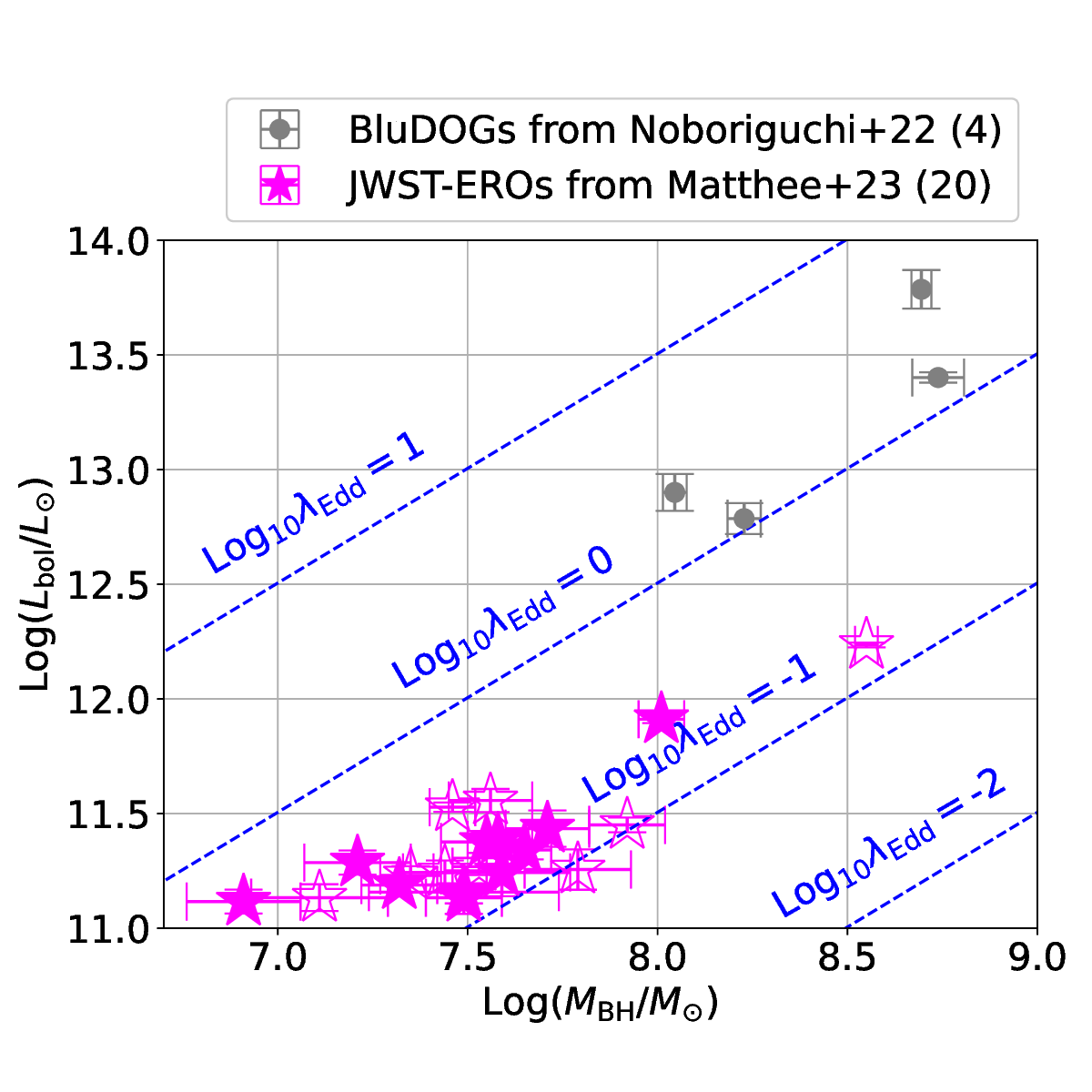}
\caption{Diagram of SMBH mass vs.\ bolometric luminosity. The gray and magenta plots denote BluDOGs from \cite{2022ApJ...941..195N} and broad-line H$\alpha$ emitters from \cite{2023arXiv230605448M}, respectively.  The blue-dashed lines represent a constant Eddington ratio of $\lambda_{\rm Edd}$ = 0.01, 0.1, 1.0, and 10.0. 
\label{fig:MBH_Lbol}}
\end{figure}

To compare the physical parameters of JWST-EROs and BluDOGs, we examine their distribution in the diagram of $M_{\rm BH}$ vs.\ $L_{\rm bol}$ (Figure~\ref{fig:MBH_Lbol}).
Note that $M_{\rm BH}$ and $L_{\rm bol}$ are estimated in different ways for JWST-EROs and BluDOGs explained in Section~\ref{sec:sample}.
Given the dusty nature of JWST-EROs, the H$\alpha$ based estimations of \cite{2023arXiv230605448M} are likely to suffer from dust extinction and should be regarded as lower limits.
As shown in Figure~\ref{fig:MBH_Lbol}, JWST-EROs and BluDOGs have log$_{10}(M_{\rm BH}/M_\odot)$ of 7--8 and 8--9, respectively, indicating that the JWST-ERO systems are one order of magnitude smaller than the BluDOG systems.
The bolometric luminosities exhibit two orders of magnitude difference as log$_{10}(L_{\rm bol}/L_\odot)$ of 11--11.5 for JWST-EROs and 12.5--14.0 for BluDOGs.
If dust extinction correction is included in the JWST-ERO estimations, these differences should become smaller. For the Eddington ratio, in which the effect of dust extinction is cancelled, 
we observe an order of magnitude difference in the Eddington ratios of JWST-EROs and BluDOGs as $\sim0.2$ and $>1$, respectively.
The Eddington ratios of JWST-EROs are similar to those of blue (normal) luminous QSOs at $z_{\rm spec} < 5$ (log$_{10}(M_{\rm BH}/M_\odot) =$8--10, log$_{10}(L_{\rm bol}/L_\odot) =$12.5--14.0: \citealt{2011ApJS..194...45S}).
Therefore, the BluDOGs are more actively accreting supermassive blackhole (SMBH) systems than JWST-EROs and normal QSOs.
Conversely, JWST-EROs appear to be a more common type of AGN than BluDOGs.

\section{Discussion} \label{sec:sd}

\subsection{Number densities of JWST-EROs and BluDOGs}\label{ss:num_dens}

The number densities of JWST-EROs and broad-line H$\alpha$ emitters are reported as $\sim4\times10^{-5}$ comoving Mpc$^{-3}$ \citep{2023arXiv230514418B} and $\sim1\times10^{-5}$ comoving Mpc$^{-3}$ \citep{2023arXiv230605448M}, respectively.
These densities are similar to those of faint X-ray AGNs reported by \cite{2015A&A...578A..83G,2019ApJ...884...19G}.
BluDOGs are significantly rarer than JWST-EROs: only 8 of 571 DOGs were found in a survey area of 105 deg$^2$ \citep{2019ApJ...876..132N}, corresponding to a number density of $\sim7\times10^{-9}$ comoving Mpc$^{-3}$ if we assume a redshift range of $2<z<3$.
The number density of DOGs is $\sim3\times10^{-7}$ comoving Mpc$^{-3}$ \citep{2015PASJ...67...86T,2019ApJ...876..132N}
There is a population of dust-obscured AGNs $\sim10$ times brighter than DOGs, called HotDOGs \citep{2012ApJ...756...96W,2012ApJ...755..173E,2015ApJ...804...27A}.
Its number density seems significantly smaller, $\sim1000$ in all sky, corresponding to $>2$ orders of magnitude smaller number density than DOGs, $\sim10^{-9}$ comoving Mpc$^{-3}$ if we assume a redshift range of $1<z<4$ \citep{2012ApJ...755..173E,2015ApJ...804...27A}.

The difference in the number densities is reasonable if we consider the difference in the luminosities and SMBH masses of these populations of AGNs.
JWST-EROs are $\sim10$ times fainter luminosities and smaller SMBH masses than BluDOGs/DOGs.
Conversely, HotDOGs are $\sim10$ times more luminous and have larger SMBH masses than BluDOGs/DOGs.
However, it is difficult to quantitatively compare their number densities as a function of luminosity or SMBH mass further because these populations have different selection methods and redshift ranges.
It would be highly interesting to develop a homogeneous selection method for these AGNs, to construct a statistical sample of them across cosmic time, and to discuss their evolution.

\subsection{Blue-Excess Fraction\label{ss:blue-excess}}

The fraction of objects with excess UV emission varies significantly between dust-obscured AGN populations. We call this the blue-excess fraction.
Although \cite{2023arXiv230514418B} only applied a single color criterion of $F277W-F444W>1.5$, the 37 selected objects have a flat UV color of $-0.5<F150W-F200W<0.5$.
Thus, their JWST-EROs have a blue-excess fraction of 100\%.
\cite{2023arXiv230605448M} constructed broad-line H$\alpha$ emitters without any color selection.
As depicted in Figure~\ref{fig:beta}, among their 20 H$\alpha$ emitters, we have identified 17 objects with $\beta_{\rm opt}>-1$, i.e., objects as red in the rest-frame optical as JWST-EROs of \cite{2023arXiv230514418B}.
11 of 17 objects have blue UV slopes as $\beta_{\rm UV}<-1.5$, which are as blue in the rest-frame UV as JWST-EROs and BluDOGs.
Therefore, the blue-excess fraction is $\sim2/3$.
Conversely, \cite{2019ApJ...876..132N} identified only 8 BluDOGs among 571 DOGs, resulting in the blue-excess fraction of $\sim1$\%.
\cite{2016ApJ...819..111A} reported the blue-excess fraction of $\sim8$\% among HotDOGs with $W4<7.4$ mag (Vega) even though it is still uncertain because of the complex selection function.
They also noted that the fraction can be smaller for their entire HotDOG sample because they have found only 2 blue-excess HotDOGs with $W4>7.4$ mag (Vega).

The significantly large blue-excess fraction ($\sim2/3$) in the broad-line H$\alpha$ emitters of \cite{2023arXiv230605448M} can be explained by selecting H$\alpha$ emission.
The dustiness of the objects is limited to the level of observability at the optical wavelength.
However, the selection by \cite{2023arXiv230514418B} was only the red optical color, and the 100\% blue-excess fraction was striking.
Conversely, DOGs/HotDOGs are selected by their extremely red color due to MIR excess emission, and these AGNs can be more dust-rich, resulting in smaller blue-excess fractions.
However, because of the different sample selections and blue-excess criteria, comparing the blue-excess fraction is complicated.
It is crucial to examine the blue-excess fraction under homogeneous sample selection and the criterion of the blue-excess in the future.

Quantifying the blue-excess fraction enables us to discuss not only the physical origin of the blue-excess emission but also the structure of the AGN core.
For example, \cite{2022ApJ...934..101A} reported a high linear polarization degree of 10.8\% in the excess UV emission from a HotDOG W0116$-$0505, demonstrating that the blue-excess is produced by scattering.
That is, the observer's line of sight to the central nucleus is through the dusty torus, the so-called Type-2 line of sight, but the UV radiation from the accretion disk and broad-line regions is scattered by something above the torus and reaches the observer.
In this case, the blue-excess fraction is related to the opening angle of the dusty torus and the covering fraction of the scattering media.
Alternatively, the dusty medium can completely cover the nucleus, but there are holes through which the blue-excess emission passes.
Scattering on the walls of the holes may also occur.
In this case, the blue-excess fraction is related to the covering fraction of the holes.

\subsection{Average UV Spectrum of BluDOGs\label{ss:exp_spec}}

\begin{figure*}
\includegraphics[width=18cm]{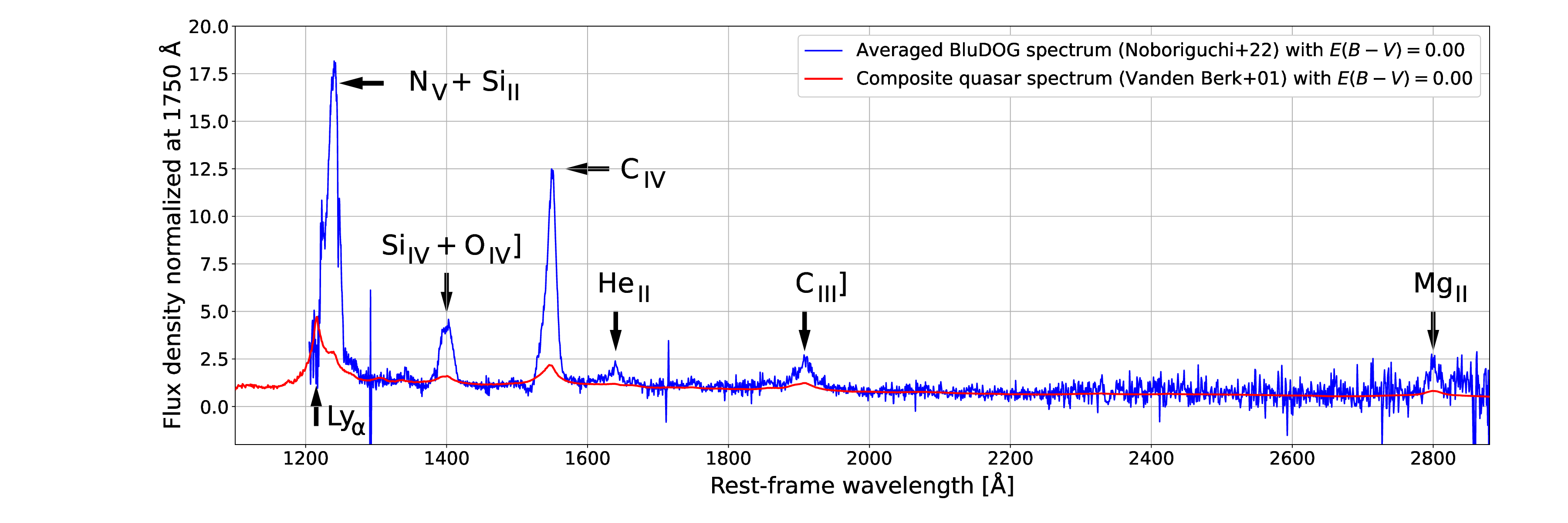}
\caption{Average spectrum of four BluDOGs \citep{2022ApJ...941..195N}. The blue line represents the dust-corrected average spectrum of BluDOGs (i.e., $E(B-V)=0.00$), and the red line represents the composite quasar spectrum \citep{2001AJ....122..549V}. The spectra are normalized by the flux density at 1750 \AA. Black arrows and text denote the detected major emission lines. The average BluDOG spectra are available online.
\label{fig:blu_ave_sp}}
\end{figure*}

The remarkable similarity between JWST-EROs and BluDOGs suggests the usefulness of an average spectrum of BluDOGs as a possible template UV spectrum for JWST-EROs.
We show average rest-UV spectra of BluDOGs based on the four spectra reported by \cite{2022ApJ...941..195N} in Figure~\ref{fig:blu_ave_sp}.
After correcting the BluDOG spectra for dust reddening $E(B-V)$ and normalizing them using the flux density at 1750 \AA, we have calculated the average spectrum.
When we correct the BluDOG spectra, we have adopted the Calzetti law \citep{2000ApJ...533..682C}.
As the sample number is restricted to four, the short/long wavelength parts are based only on one or two objects.\footnote{Each spectral section is based on the objects listed as follows:
$\lambda_{\rm rest} =$ 1210--1360 \AA\ based on J1443,
$\lambda_{\rm rest} =$ 1360--1490 \AA\ based on J1202 and J1443,
$\lambda_{\rm rest} =$ 1490--1850 \AA\ based on J0907, J1202, J1207, and J1443,
$\lambda_{\rm rest} =$ 1850--2080 \AA\ based on J0907, J1202, and J1207,
$\lambda_{\rm rest} =$ 2080--2270 \AA\ based on J0907 and J1207, and
$\lambda_{\rm rest} =$ 2270--2880 \AA\ based on J0907.}
In Figure~\ref{fig:blu_ave_sp}, we also show the quasar average spectra \citep{2001AJ....122..549V} without dust reddening as references.

The BluDOG spectrum exhibits extremely strong emission lines compared with the average spectrum of quasars.
The rest-frame EWs of the C~{\sc iv} line of BluDOGs range from 100 to 200 \AA\ \citep{2022ApJ...941..195N}, whereas the SDSS quasars in the similar luminosity range show EWs well smaller than 100 \AA\ \citep{2011ApJS..194...45S}.
The C~{\sc iv} EWs are 2--$4\sigma$ above the C~{\sc iv} EW distribution of SDSS quasars.
Such large EWs are also observed in the blue-excess HotDOG W0116$-$0505 reported by \cite{2020ApJ...897..112A}.
Although the physical origin of the large EWs is still unclear, a possible scenario is different dust attenuation amounts for the accretion disk (i.e. continuum) and the broad-line regions (i.e. emission lines).
Namely, the accretion disk is more heavily obscured than the broad-line regions, leading to the fainter continuum and the larger emission line EWs.
As an extreme case of this scenario, we may consider that only the continuum suffers from dust reddening but the emission lines do not.
Then, if we apply dust correction only for the continuum, the C~{\sc iv} EWs are reduced to $\approx60$ \AA, which is consistent with the average value of the SDSS quasars for the luminosity range.
This is indicative, albeit an extreme case.


We may expect the spectrum of the rest-UV blue-excess of JWST-EROs to be similar to that of BluDOGs, which is composed of a moderately reddened continuum from the AGN accretion disk and broad emission lines with extremely large EWs.
However, \cite{2023arXiv230805735F} and \cite{2023arXiv230811610K} showed the spectra of two JWST-ERO-like objects without strong metal emission lines in the rest-frame UV.
\cite{2023arXiv230810900L} reported the UV to optical spectra of some of the JWST-ERO sample of \cite{2023arXiv230607320L}, showing no strong metal emission lines in UV, either, in addition to a report of three brown dwarfs in the Milky Way as contaminants
(see also \citealt{2023arXiv230812107B}).
Although a systematic and statistical sample of the UV spectra of JWST-EROs is required for a firm conclusion in the future, these initial results show a difference in the UV metal emission line strength between JWST-EROs and BluDOGs, implying differences in the physical properties of the AGN systems, such as metallicity.

\subsection{An Attempt to Locate the Dusty AGN Populations in a Coherent Picture}

\begin{figure*}
\includegraphics[width=18cm]{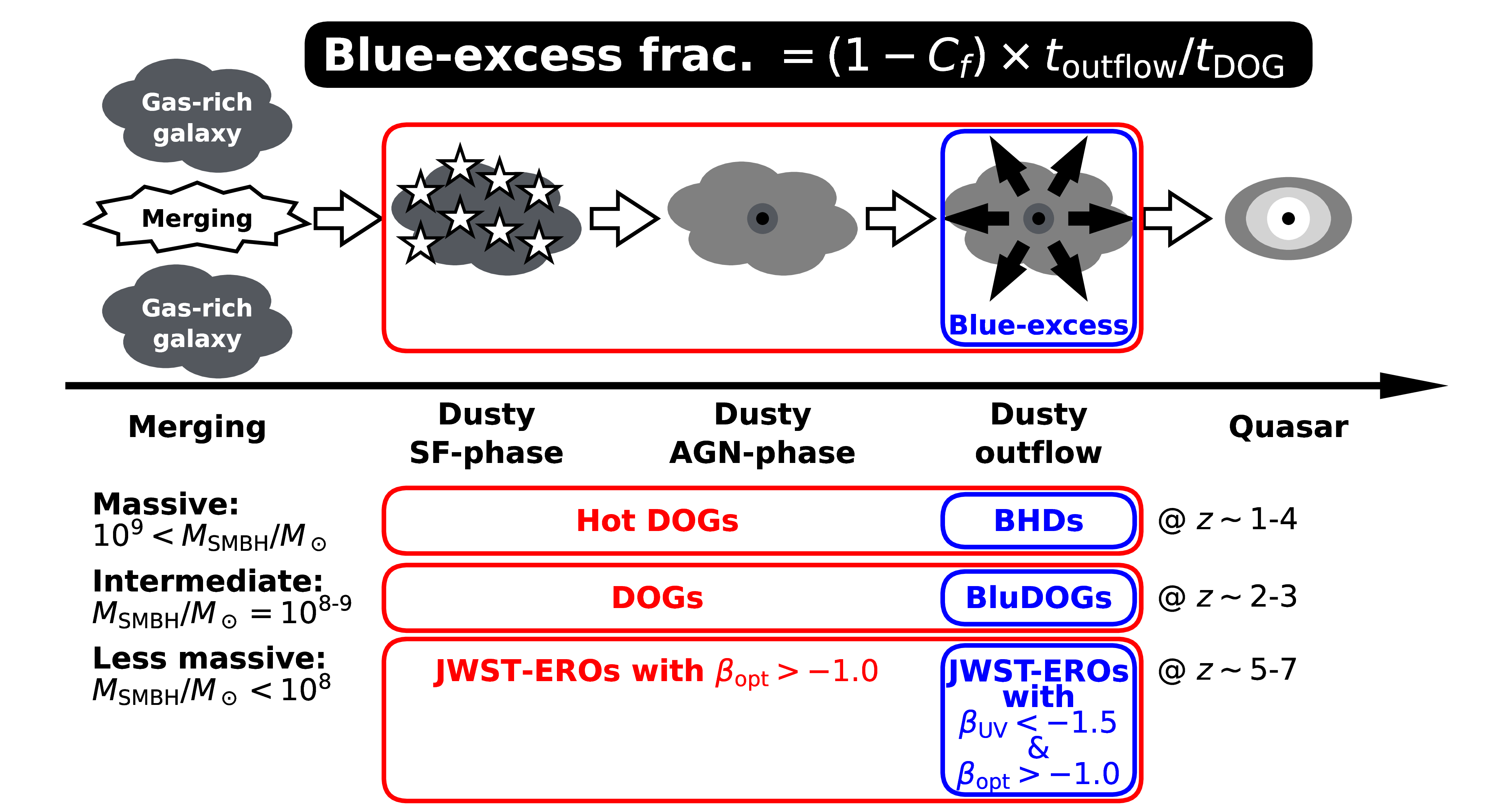}
\caption{Dusty AGN populations in the context of gas-rich major merger scenario for AGN formation.
The blue-excess is observable through less-dusty channels produced by outflows.
Therefore, its fraction is determined by the covering fraction of dusty media ($C_f$) and the ratio of the time-scales of the outflow phase ($t_{\rm outflow}$) to the dust obscured galaxy phase ($t_{\rm DOG}$). 
\label{fig:evo}}
\end{figure*}

We have discussed three dusty AGN populations of JWST-EROs, DOGs, and HotDOGs and their blue-excess emissions in the rest-frame UV range.
Here, we attempt to consider the relation between these dusty AGN populations.
As shown in Figure~\ref{fig:MBH_Lbol}, the SMBH masses in JWST-EROs is log$_{10}(M_{\rm BH}/M_\odot)\sim7.5$, and those in BluDOGs is log$_{10}(M_{\rm BH}/M_\odot)\sim8.5$.
The SMBH masses of HotDOGs are close to or greater than log$_{10}(M_{\rm BH}/M_\odot)\sim9$ \citep{2018ApJ...852...96W}.
Thus, these dusty AGN populations are different on the SMBH mass scale.
However, the overall SEDs of JWST-EROs are similar to those of BluDOGs and blue-excess HotDOGs, indicating physical similarity.
Therefore, we propose a unified picture for the three dusty AGN populations in the gas-rich major merger scenario of AGN formation \citep{2006ApJS..163....1H}, as shown in Figure~\ref{fig:evo}.

A major merger event triggers intense star formation that is heavily obscured by the dusty interstellar medium (Dusty SF-phase).
For example, gas feeding onto the central SMBH can occur because of the radiation drag effect during a strong dusty starburst \citep{2001ApJ...560L..29U}, which ignites the AGN still obscured by the surrounding dusty medium (Dusty AGN-phase).
As a feedback effect by the AGN, the outflow partly clears the dusty medium and a part of the UV emission escapes through the medium, producing the blue-excess emission (Dusty outflow phase).
Finally, a significant part of the dusty medium is cleared, and the central AGN can be observed as an unobscured quasar (Quasar phase).
In this scenario, dusty AGNs with blue-excess such as BluDOGs and JWST-EROs with $\beta_{\rm UV}<-1.5$ are observed in the Dusty outflow phase.\footnote{As described in Section~\ref{sec:intro}, BluDOGs exhibit an evidence of outflows in their spectra \citep{2022ApJ...941..195N}. It is unclear whether DOGs without blue-excess exhibit outflows or not due to the lack of spectroscopic information. However, if outflows produce less-dusty channels, the blue-excess would be observed through the channels. Therefore, in Figure~\ref{fig:evo}, we simply consider that the blue-excess is only observable in the outflow phase.}

As discussed in Section \ref{ss:blue-excess}, the blue-excess fraction can differ significantly between the three dusty AGN populations, although the selection method and criterion of the blue-excess are not yet homogeneously defined.
The different blue-excess fraction implies differences in the evolutionary time scale, outflow efficiency, or the geometry of the dusty medium around the nucleus, as discussed in Section \ref{ss:blue-excess}.
In Figure~\ref{fig:evo}, we note an equation describing the blue-excess fraction, which is summarized as a combination of the covering fraction of dusty media and the time-scales of the outflow and DOG phases.

If the JWST-EROs are in the Dusty outflow phase, there can be a significant number of galaxies in the Dusty SF and AGN phases without blue-excess in the epoch of cosmic dawn.
However, the current estimate of the blue-excess fraction in JWST-EROs is $\sim1$, implying that there is not much room for the dusty SF/AGN phases.
The JWST-EROs are the AGN population in the very early Universe such as $z>5$.
Possible lower metallicity and a smaller dust-to-gas ratio in the interstellar medium may reduce the covering fraction of the dusty media and/or shorten the time scale of the DOG phase, resulting in a larger blue-excess fraction.
However, the time scale of the DOG phase remains uncertain.
\cite{2022ApJ...936..118Y} suggested the entire time scale to be 4 Myrs from their SED simulation of DOGs based on the post-processed radiation transfer calculation after hydrodynamic simulations of a major merger event.
It will be interesting to examine any metallicity dependence on the time scale in future work.

Less massive AGNs similar to JWST-EROs should exist in cosmic noon when DOGs/HotDOGs are observed.
These less massive AGNs are missing thus far because of the limited sensitivity of the MIR survey data based on WISE and the limited survey areas of Spitzer/MIPS 24 $\mu$m imaging (references).
For example, JWST-EROs are 100 times fainter than BluDOGs in bolometric luminosity, corresponding to a 5 mag difference.
JWST MIRI survey data would be useful to search for the cosmic noon counterpart of high-$z$ JWST-EROs, but the limited survey area would be a bottleneck.
For wide-field imaging surveys, the Euclid and Roman Space Telescope will emerge in the near future.
However, their wavelength coverage is limited to the wavelength $<2$ $\mu$m, which would not be sufficiently long to sample JWST-ERO-like objects.
A Japanese future space mission concept, GREX-PLUS \citep{2022SPIE12180E..1II,2023arXiv230408104G} will conduct a 1,000 deg$^2$ imaging survey with wavelength coverage up to 8 $\mu$m.
Because the GREX-PLUS survey depths are $>5$ mag deeper than those of {\textit{WISE}} used in \cite{2019ApJ...876..132N}, a significant number of DOGs, including optically dark AGNs, \citep[e.g.,][]{2020ApJ...899...35T} as faint as JWST-EROs would be detected.

\begin{acknowledgments}

The authors gratefully acknowledge the anonymous referee for a careful reading of the manuscript and very helpful comments.

This work was supported by JSPS KAKENHI Grant Numbers JP23H00131 (A.K.I.), JP23H01215 (T.N.), JP22H01266 (Y.T.), and 21H01126 (T.M.).
\end{acknowledgments}

\vspace{5mm}

\software{astropy \citep{2013A&A...558A..33A,2018AJ....156..123A,2022ApJ...935..167A}}

\bibliography{Ref}{}
\bibliographystyle{aasjournal}

\end{document}